\documentclass[11pt, reqno]{amsart}
\usepackage[foot]{amsaddr}
\usepackage{indentfirst}
\usepackage{fullpage}
\usepackage{latexsym}
\usepackage{amsmath}
\usepackage{mathtools}
\usepackage{eqparbox}
\usepackage{algorithm}
\usepackage{algpseudocode}
\usepackage{hyperref}
\usepackage{float}
\usepackage{cite}
\usepackage{nicefrac}
\usepackage{array}
\usepackage{multirow}
\usepackage{hhline}
\usepackage{graphicx}
\usepackage[toc,page]{appendix}
\usepackage{verbatim}
\usepackage{enumerate}

\title{BoolSi: a tool for distributed simulations and analysis of Boolean networks}
\author{Vladyslav Oles$^1$}
\address{$^1$Department of Mathematics and Statistics, Washington State University}
\email{vladyslav.oles@wsu.edu}
\author{Anton Kukushkin}
\address{}
\email{kukushkin.anton@gmail.com}

\begin{document}
\maketitle
\begin{abstract}
	We present BoolSi, an open-source cross-platform command line tool for distributed simulations of deterministic Boolean networks with synchronous update. It uses MPI standard to support execution on computational clusters, as well as parallel processing on a single computer. BoolSi can be used to model the behavior of complex dynamic networks, such as gene regulatory networks. In particular, it allows for identification and statistical analysis of network attractors. We perform a case study of the activity of a cambium cell to demonstrate the capabilities of the tool.
\end{abstract}
\section{Introduction}
\subsection{Background}
A Boolean network (BN) is a discrete dynamical system consisting of nodes that represent objects from application domain, and their update rules, that represent interactions between the objects. Update rule of a node is a Boolean function that defines the next state of the node based on current states of nodes. Therefore, each node can take one of two possible states, denoted 0 or 1. In synchronous BNs, the next state of the network is calculated by applying all update rules simultaneously. In asynchronous BNs, the states of only a subset of network nodes get updated at a time. The exact nodes to update at each time step can be either determined by some pre-selected rule, or chosen at random, which introduces non-determinism into system dynamics.

State space of a BN is finite and consists of $2^n$ network states, where $n$ is the number of nodes in the network. Simulating a BN with deterministic dynamics will eventually cause the network to reach a previously visited state and thus fall into a cycle called an attractor. In the case of a cycle of length 1, the attractor is called a steady state. States of the network that lead to a particular attractor constitute its basin of attraction. Attractors represent long-term behavior of the modeled processes and may therefore carry important information about them \cite{kauffman93, li04}. 

Using BNs to model continuous real-world processes requires binary interpretation of their states, which is usually achieved through thresholding. Despite such simplification, BN models can capture important properties of the dynamics of complex systems, which makes them a practical choice in a range of domains, including systems biology \cite{davidich08}, robotics \cite{roli11}, epidemiology \cite{newman02}, and economics \cite{alexander03, paczuski00}. 

\subsection{Existing software packages}
Due to the practicality of the model, a number of software tools for BN simulations have been made available.

BoolNet \cite{mussel10} is an R package for construction, simulation and analysis of synchronous, asynchronous, and random Boolean networks. BoolNet supports identification of synchronous attractors by both exhaustive and heuristic search.

BooleanNet \cite{albert08} is a Python software library for simulation and visualization of synchronous and asynchronous BNs, capable of detecting their attractors. Unlike most other tools from the field, BooleanNet allows for the simulation of simulate discrete update rules as piecewise linear differential equations by introducing continuous variables to the model.

RaBooNet \cite{ferranti17} is a Matlab toolbox for generation and simulation of random BNs, with an emphasis on their application to supervised machine learning. RaBooNet does not support user-defined update rules of a network, only allowing rules to be randomly generated based on network structure.

BioNetGen \cite{blinov04}, GINsim \cite{gonzalez06} and PlantSimLab \cite{ha19} are software packages with graphical user interface, aimed at rule-based modeling of biochemical systems. All three packages can operate with both binary-state and multi-state models. 

\subsection{Our contribution}
Here, we present BoolSi, a command line tool for simulation and analysis of deterministic BNs with synchronous update.

The main distinguishing characteristic of BoolSi is the support for MPI standard to run simulations in parallel, e.g. on a computational cluster. To the best of our knowledge, this is the first BN modeling tool capable of high-performance computing. 

Another unique feature of BoolSi is the analysis that it performs on the identified attractors. This approach, inspired by statistical mechanics, was first introduced in \cite{oles17} and used therein to better understand the role of hormonal signals in regulating cambium proliferation.

\section{Functionality}
BoolSi requires user to define nodes, update rules, and initial states of a BN from simple text input. It supports simulating the network from a range of initial states within a single run, and aggregating the results of simulations. In particular, BoolSi can use exhaustive search to identify all network attractors, as well as their trajectory length distributions and the sizes of their basins of attraction. User can limit the length of attractors to search for --- for example, setting the limit to 1 will identify the steady states only. Another mode of operation aims to find conditions that lead to specific states of the network.

It is possible to override the simulation process, which can be used to model external influence on the network. User can fix the state of any node for the entire simulation (e.g. modeling gene knockout in regulatory networks), or perturb it at any time step (e.g. modeling sensory input in robotics).

To speed up simulations, BoolSi stores pre-evaluated update rules as truth tables. This requires $O(n 2^d)$ memory, where $n$ is the number of nodes in the network and $d$ is its maximum in-degree. Such an approach imposes a practical limitation on the in-degree of a node, with the majority of modern computers being able to handle the nodes of in-degree up to 20. This however is sufficient for many BN models from the literature, where in-degrees rarely exceed 10 (see e.g. \cite{aldana03}, \cite{bornholdt08} and the references therein).

\subsection{Attractor analysis}
\label{attractor analysis}
Boolsi performs statistical analysis of the found attractors to extract information about the interplay between the nodes. Let $n$ denote the number of nodes in a BN. Simulating the network from each of its possible initial states $\mathbf{s}^k \in \{0, 1\}^n, k = 1, \ldots, 2^n$, leads to an attractor, comprised of one or multiple network states. We define $\alpha_k^i \in [0, 1]$, the activity coefficient of node $i$ associated with network state $\mathbf{s}^k$, as the average of 0 and 1 states of the node in the attractor reachable from $\mathbf{s}^k$. All activity coefficients of node $i$ form vector $\boldsymbol\alpha^i \in [0, 1]^{2^n}$, which represents long-term activity of the node with regard to the entire state space of the network. For every pair of nodes $i$ and $j$, BoolSi computes Spearman's correlation coefficient $\rho_{ij} \in [-1, 1]$ between $\boldsymbol\alpha^i$ and $\boldsymbol\alpha^j$, interpreted as the degree of mutual amplification of the two nodes. We note that the impact of each attractor on $\rho_{ij}$ is proportional to the size of its basin of attraction. Spearman's correlation measures the strength and direction of a monotonic association between two variables, and was chosen for its ability to capture non-linear relationships.

\begin{figure}[H]
	\includegraphics[width=\linewidth]{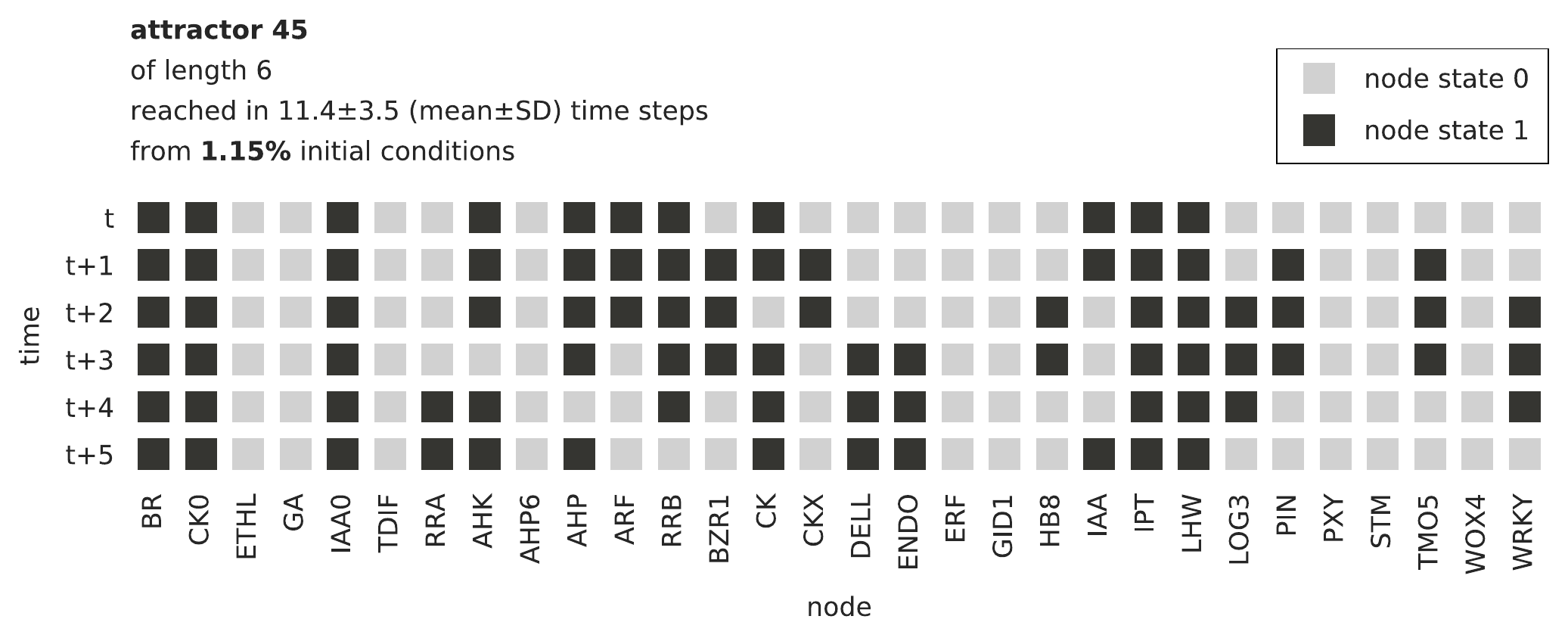}
	\caption{BoolSi output showing an attractor of a 30-node network (model of cambium regulation from \cite{oles17}, used in Section \ref{case study}). The activity coefficient of the node CK associated with each of roughly $0.0115 \cdot 2^{30} \approx 12,348,031$ network states $\mathbf{s}^k$ leading to this attractor is $\alpha_k^{\mathrm{CK}} = \frac{5}{6}$.}
	\label{fig:attractor}
\end{figure}

\begin{figure}[H]
	\includegraphics[width=\linewidth]{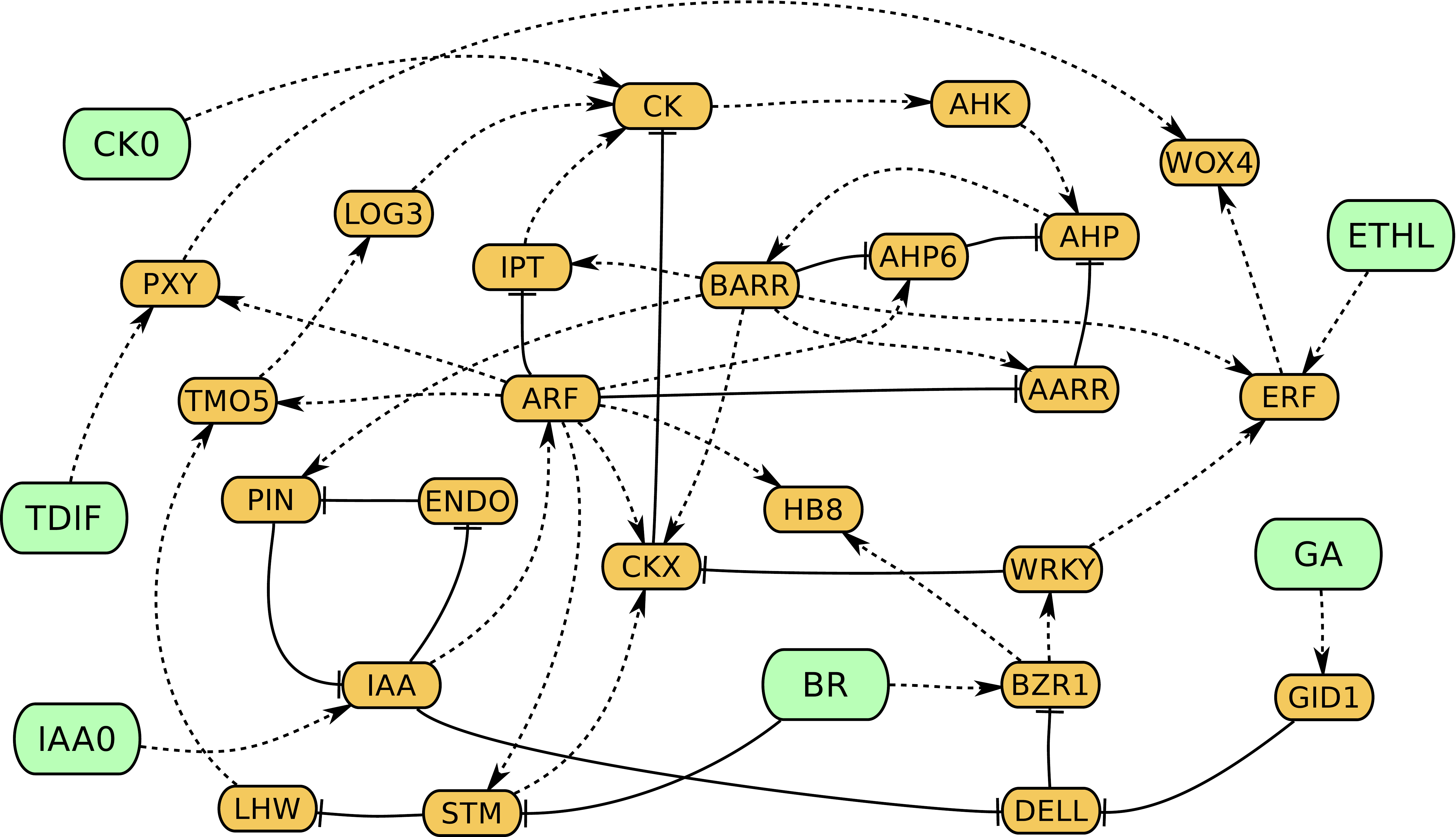}
	\vspace{.1in}
	\caption{Cambium regulation network from \cite{oles17}. Bigger, green-colored nodes represent hormonal signals produced outside cambium; their states do not change throughout each simulation run. Dashed and solid lines represent activation and inhibition respectively.}
	\label{fig:model}
\end{figure}

\subsection{User interface}
BoolSi provides a command line interface. It relies on YAML input file to define nodes and update rules of a BN and to provide simulation details, such as initial states. Update rules are described using common syntax for logical functions, including operators "and", "or", "not".

\subsection{Output formats}
BoolSi supports a number of output formats: PDF documents, vector or raster image files (SVG, PNG, TIFF), or CSV text files to allow for machine processing.

\section{Case study}
\label{case study}
Cambium is a plant tissue layer that produces cells for plant growth. It is concealed under multiple layers of other cells, which hinders identification of mutants with altered secondary growth and isolation of live cambium cells. As a consequence, understanding the biology of cambium remains incomplete. Mathematical modeling has the potential to help with predicting the outcome of interactions controlling cambium activity \cite{oles17}.

We based the case study off of the 30-node BN model of cambium regulation (see Figure \ref{fig:model}) described in \cite{oles17}, formatted as a YAML input file for BoolSi (see Appendix \ref{cambium}). We used BoolSi to find network attractors by simulating from all $2^{30} \approx 1.07 \cdot 10^9$ initial states of the network, and to analyze the found attractors. The exhaustive search took about 45 hours on a cluster of 21 computers with 8 CPU cores each. Of the total 168 CPU cores, 112 were 4 GHz, 17 were 3.4 GHz, 11 were 2.1 GHz, and 28 were 1.4 GHz.

\begin{figure}[H]
	\includegraphics[width=\linewidth]{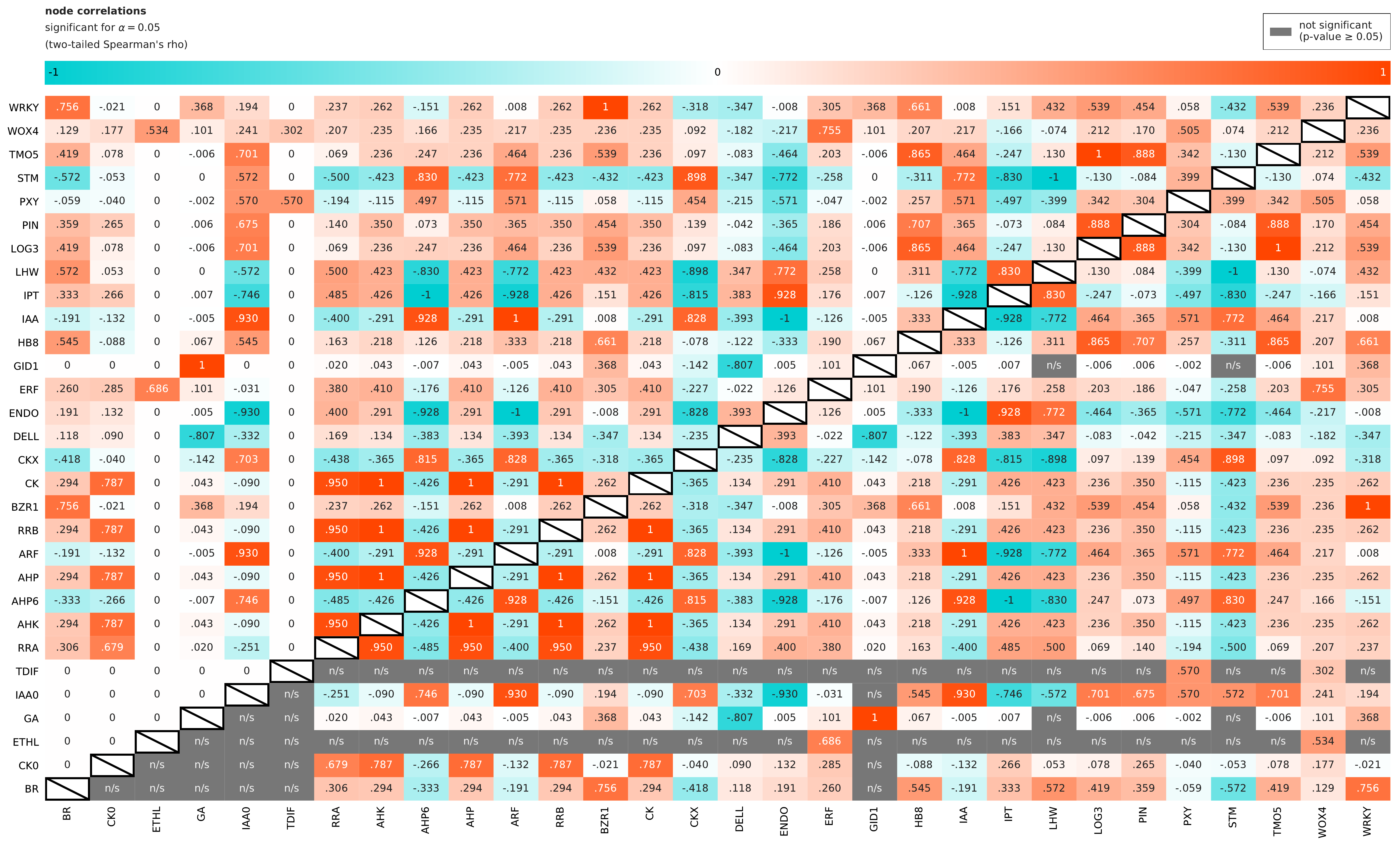}
	\vspace{.04in}
	\caption{BoolSi output showing the results of attractor analysis. Cell in $i$-th row and $j$-th column shows $\rho_{ij}$, Spearman's correlation coefficient between activity of the nodes $i$ and $j$ in the attractors. In the lower-right triangle, the values corresponding to p-values $ \geq 0.05$ are hidden.}
	\label{fig:node correlations}
\end{figure}

The results of attractor analysis (Figure \ref{fig:node correlations}) agree with the findings of \cite{oles17} and experimental data. In particular, each of the nodes IAA and BR showed positive correlation with both WOX4 and HB8, whose combined activity represents cambium proliferation. Node ETHL exhibited strong positive correlation ($> 0.5$) with WOX4, but no significant correlation with HB8. The analysis also reproduced the mutually inhibitory relationship between cytokinin and auxin signaling \cite{schaller15}, manifested as the negative correlation between the nodes IAA and CK. Unlike in \cite{oles17}, our analysis found a positive effect of the nodes CK and TDIF on the expression of WOX4, as well as a positive relationship between CK and HB8 (the correlation between TDIF and HB8 was 0). This is consistent with published experimental data on the key role of both CK and TDIF in regulation of cambium activity. We note that BoolSi automatically extracts information about the interplay between each pair of nodes, stepping up the approach of manually inferring the relationships between the selected nodes.

\subsection{Testing the effects of mutations}
To test if BoolSi can accurately model the mutations of a cambium cell, we simulated the effect of mutations known to reduce cambium proliferation: double knockout of ERFs \textit{erf108erf109}, a quadruple knockout of IPT which was deficient in cytokinin synthesis, and mutants in cytokinin receptor AHK4 \cite{oles17}. Mutations were mimicked by fixing the states of ERF, IPT, or AHK to be 0 in the simulations (see Appendix \ref{mutants} for the corresponding BoolSi input). In agreement with the experimental data, all mutations resulted in lower proliferation activity, measured as $\displaystyle \sqrt{\frac{||\boldsymbol\alpha^{\mathrm{WOX4}}||_1^2 + ||\boldsymbol\alpha^{\mathrm{HB8}}||_1^2}{2}}$ (see Subsection \ref{attractor analysis}). The proliferation activity in \textit{erf}, \textit{ipt}, and \textit{ahk} was, respectively, 1.91, 1.16, and 1.02 times lower than in the model of cambium unaffected by mutations. We note that the insignificance of the decrease of proliferation activity in \textit{ahk} suggests the direction for improvement of the model.

\section{Conclusion}
We presented BoolSi, an open-source tool for distributed parallel simulations of Boolean networks with synchronous update. To the best of our knowledge, this is the first BN modeling tool capable of high-performance computing. In addition, BoolSi incorporates a novel method for analyzing found attractors to extract information about the interplay between the nodes.

We used BoolSi to perform a case study of the activity of a cambium cell. The obtained results complied with both experimental data and previous findings about the model. This suggests that BoolSi can potentially speed up BN simulations and analysis, in particular those conducted in the field of systems biology. 

\section{Availability and requirements}
\noindent Project page: \url{http://github.com/boolsi/boolsi}
\\Programming language: Python 3.4 or higher
\\Operating systems: Linux, Windows, MacOS

\section{Acknowledgements}
The authors are grateful to the Department of Mathematics and Statistics of Washington State University for providing access to the computer cluster used in the case study. In addition, the authors would like to thank Kevin Cooper for his guidance on MPI standard and Oracle Grid Engine software, Alexander Panchenko for suggestions on the functionality of BoolSi, Taras Nazarov for their assistance with the biology part of the paper, Iana Khotsianivska and Aleksey Lapenko for helping with the design of BoolSi output, and Grace Grimm for reviewing and editing.

\appendix
\section{Input file for cambium cell model}
\label{cambium}
\begin{verbatim}
nodes:
    - BR
    - CK0
    - ETHL
    - GA
    - IAA0
    - TDIF
    - RRA
    - AHK
    - AHP6
    - AHP
    - ARF
    - RRB
    - BZR1
    - CK
    - CKX
    - DELL
    - ENDO
    - ERF
    - GID1
    - HB8
    - IAA
    - IPT
    - LHW
    - LOG3
    - PIN
    - PXY
    - STM
    - TMO5
    - WOX4
    - WRKY

update rules:
    BR: BR
    CK0: CK0
    ETHL: ETHL
    GA: GA
    IAA0: IAA0
    TDIF: TDIF
    RRA: RRB and not ARF
    AHK: CK
    AHP6: ARF and not RRB
    AHP: AHK and not (RRA and AHP6)
    ARF: IAA
    RRB: AHP
    BZR1: BR and not DELL
    CK: majority(CK0, IPT, LOG3, not CKX, 0)
    CKX: majority(ARF, RRB, not WRKY, STM, 0)
    DELL: not (IAA or GID1)
    ENDO: not IAA
    ERF: ETHL and (RRB or WRKY)
    GID1: GA
    HB8: ARF and BZR1
    IAA: IAA0 and not PIN
    IPT: RRB or not ARF
    LHW: not STM
    LOG3: TMO5
    PIN: RRB and not ENDO
    PXY: TDIF and ARF
    STM: ARF and not BR
    TMO5: LHW and ARF
    WOX4: PXY or ERF
    WRKY: BZR1

initial state:
    BR: any
    CK0: any
    ETHL: any
    GA: any
    IAA0: any
    TDIF: any
    RRA: any
    AHK: any
    AHP6: any
    AHP: any
    ARF: any
    RRB: any
    BZR1: any
    CK: any
    CKX: any
    DELL: any
    ENDO: any
    ERF: any
    GID1: any
    HB8: any
    IAA: any
    IPT: any
    LHW: any
    LOG3: any
    PIN: any
    PXY: any
    STM: any
    TMO5: any
    WOX4: any
    WRKY: any


\end{verbatim}
\section{Input file modifications for modeling the effects of mutations}
\label{mutants}
\begin{enumerate}[{1.}]
	\item Modifications of the input file (see Appendix \ref{cambium}) for modeling ERF mutant background:
\begin{verbatim}
83c83
<     ERF: any
---
>     ERF: 0
97c97,98
< 
---
> fixed nodes:
>     ERF: 0

\end{verbatim}
	\item Modifications of the input file for modeling IPT mutant background:
\begin{verbatim}
83c83
<     IPT: any
---
>     IPT: 0
97c97,98
< 
---
> fixed nodes:
>     IPT: 0

\end{verbatim}
	\item Modifications of the input file for modeling ERF mutant background:
\begin{verbatim}
83c83
<     AHK: any
---
>     AHK: 0
97c97,98
< 
---
> fixed nodes:
>     AHK: 0

\end{verbatim}
\end{enumerate}


\begin{thebibliography}{99}
\bibitem{albert08}
Albert, I., Thakar, J., Li, S., Zhang, R., \& Albert, R. (2008). Boolean network simulations for life scientists. \textit{Source code for biology and medicine, 3}(1), 16.

\bibitem{aldana03}
Aldana, M., Coppersmith, S., \& Kadanoff, L. P. (2003). Boolean dynamics with random couplings. In \textit{Perspectives and Problems in Nolinear Science} (pp. 23-89). Springer, New York, NY.

\bibitem{alexander03}
Alexander, J. M. (2003). Random Boolean networks and evolutionary game theory. \textit{Philosophy of Science, 70}(5), 1289-1304.

\bibitem{blinov04}
Blinov, M. L., Faeder, J. R., Goldstein, B., \& Hlavacek, W. S. (2004). BioNetGen: software for rule-based modeling of signal transduction based on the interactions of molecular domains. \textit{Bioinformatics, 20}(17), 3289-3291.

\bibitem{bornholdt08}
Bornholdt, S. (2008). Boolean network models of cellular regulation: prospects and limitations. \textit{Journal of the Royal Society Interface, 5}(suppl\_1), S85-S94.

\bibitem{davidich08}
Davidich, M. I., \& Bornholdt, S. (2008). Boolean network model predicts cell cycle sequence of fission yeast. \textit{PloS one, 3}(2), e1672.

\bibitem{ferranti17}
Ferranti, D., Krane, D., \& Craft, D. (2017). The value of prior knowledge in machine learning of complex network systems. \textit{Bioinformatics, 33}(22), 3610-3618.

\bibitem{gonzalez06}
Gonzalez, A. G., Naldi, A., Sanchez, L., Thieffry, D., \& Chaouiya, C. (2006). GINsim: a software suite for the qualitative modelling, simulation and analysis of regulatory networks. \textit{Biosystems, 84}(2), 91-100.

\bibitem{ha19}
Ha, S., Dimitrova, E., Altarawy, D., Hosamadine, H., Deb, D., Hilmer, R., ... \& Glazebrook, J. (2019). PlantSimLab-A Modeling and Simulation Web Tool for Plant Biologists. \textit{BMC Bioinformatics, Accepted}.

\bibitem{kauffman93}
Kauffman, S. A. (1993). \textit{The origins of order: Self-organization and selection in evolution}. OUP USA.

\bibitem{li04}
Li, F., Long, T., Lu, Y., Ouyang, Q., \& Tang, C. (2004). The yeast cell-cycle network is robustly designed. \textit{Proceedings of the National Academy of Sciences, 101}(14), 4781-4786.

\bibitem{mussel10}
M\"ussel, C., Hopfensitz, M., \& Kestler, H. A. (2010). BoolNet—an R package for generation, reconstruction and analysis of Boolean networks. \textit{Bioinformatics, 26}(10), 1378-1380.

\bibitem{newman02}
Newman, M. E. (2002). Spread of epidemic disease on networks. \textit{Physical review E, 66}(1), 016128.

\bibitem{oles17}
Oles, V., Panchenko, A., \& Smertenko, A. (2017). Modeling hormonal control of cambium proliferation. \textit{PloS one, 12}(2), e0171927.

\bibitem{paczuski00}
Paczuski, M., Bassler, K. E., \& Corral, Á. (2000). Self-organized networks of competing boolean agents. \textit{Physical Review Letters, 84}(14), 3185.

\bibitem{roli11}
Roli, A., Manfroni, M., Pinciroli, C., \& Birattari, M. (2011, April). On the design of Boolean network robots. In \textit{European Conference on the Applications of Evolutionary Computation} (pp. 43-52). Springer, Berlin, Heidelberg.

\bibitem{schaller15}
Schaller, G. E., Bishopp, A., \& Kieber, J. J. (2015). The yin-yang of hormones: cytokinin and auxin interactions in plant development. \textit{The Plant Cell, 27}(1), 44-63.
\end{thebibliography}
\end{document}